\documentclass{aa}
\usepackage{graphicx}
\usepackage{txfonts}
\begin{document}

\title{Estimation of time delays from unresolved photometry}

   \author{
          A. Hirv
          \and
          T. Eenm\"ae
          \and
          T. Liimets
          \and
          L. J. Liivam\"agi 
          \and
          J. Pelt
    }

   \offprints{A. Hirv}
   \authorrunning{A. Hirv et al.}
   \titlerunning{Estimation of time delays}	
   \institute{
             Tartu Observatory, 61602 T\~{o}ravere, Estonia
             }

   \date{Received ---; accepted ---}

   \abstract
   {Longtime monitoring of gravitational lens systems is often done using 
telescopes and recording equipment with modest resolution. Still, it would be 
interesting to get as much information as possible from the measured 
lightcurves.
From 
high resolution images we know that the recorded quasar images are often blends 
and that the corresponding time series are not pure shifted replicas of the
source 
variability. }
   {In this paper we will develop an algorithm to unscramble this kind of
blended data.}
   {The proposed method is based on a simple idea. We use one of the
photometric 
curves, which is supposedly a simple shifted replica of the source curve, 
to
build 
different artificial combined curves. Then we compare these artificial curves
with the blended curves. Proper solutions for a full set of time delays are then obtained
by varying free input parameters and estimating statistical distances between 
the artificial and blended curves.}
   {We performed a check of feasibility and applicability of the new algorithm. 
For numerically generated data sets the time delay systems were recovered for a 
wide range of setups. Application of the new algorithm to the classical double 
quasar QSO 0957+561 A,B lightcurves shows a clear splitting of one of the
images. 
This is an unexpected result and extremely interesting, especially in the context
of 
the recent controversy about the exact time delay value for the system.}
   {The proposed method allows us to properly analyse the data from low 
resolution 
observations that have long time coverages. There are a number of 
gravitational
lens 
monitoring programmes that can make use of the new algorithm.}
          

   \maketitle

\section{Introduction} 

Gravitationally lensed quasar images can be monitored for a long period of 
time. The obtained time series can then be used to estimate time delays 
for different light paths to the observer. The best example of this kind 
of long-term photometry is a time series obtained by R. Schild and 
collaborators (see for instance Schild et al.~\cite{Schild}) for the 
ubiquitous system QSO 0957+561. Typically, such long programmes can be 
carried out only on small telescopes and consequently they will have a 
modest resolution. For many interesting multiply lensed systems, some of 
the images may remain unresolved. It occurs that for a reasonably long 
time series it is possible to recover complex time delay systems even from 
the blended images.

There are a number of different time delay estimation methods used by various
research 
groups. For a short review of the popular methods see for instance Kundi\`c et 
al.~(\cite{Kundic}). Some recent more peculiar approaches can be found in 
Hjorth~(\cite{Hjorth}), Pijpers~(\cite{Pijpers}), Barkana~(\cite{Barkana}), 
Burud~(\cite{Burud}), and Gil-Merino~(\cite{GilMerino}). These methods 
can roughly be divided into three classes: 
\begin{itemize} 
\item cross-correlation based methods; 
\item methods based on interpolation (linear, polynomial, spline, etc.); 
\item methods which 
use dispersion spectra. 
\end{itemize} 
In principle, nearly all the known methods can be 
reformulated so that they can handle more complex models of time series, 
such as those treated in 
the current paper. Because of the simplicity and familiarity of the 
last class of methods, the new methods described hereafter are also based 
on the computing of dispersion
spectra, 
introduced in Pelt et al. (\cite{Pelt94}) and refined in Pelt et al.
(\cite{Pelt96}).

In Pelt et al. (\cite{Pelt98}), dispersion analysis was generalised for systems
with 
multiple images. However, in previous papers and algorithms it was always 
assumed that observed curves are time-shifted replicas (perhaps distorted
by 
microlensing) of the source variability. In practice, especially if we deal
with 
low resolution photometry, this is not always the case. Often 
the photometric aperture 
covers several weak images and what we effectively get is the blend of 
different lightcurves. From physical considerations we can predict that 
the components of the blends are certain weighted sums of the time-shifted 
source 
variability curves. Typically, the time delays between blended components are 
remarkably shorter than the delays between images, which are significantly separated.
A singular case of blending, where we observe only the sum of multiple 
components, was analysed in Geiger~(\cite{Geiger}). In this method a certain
amount of extra information is used from additional interferometric
observations.

Our paper is organised as follows. First we introduce general ideas of the
proposed 
new algorithms. Then we formulate the algorithms in the terms of the 
specific time
series 
operations involved. In the next part we describe the numerical tests
we made to 
evaluate the algorithms. Then we consider applicability of the method in 
different 
contexts and apply the method to a real observed data sequence -- 
the classical 
system
QSO 
0957+561 A,B. It is well known (from radio observations and deep images) 
that the
photometric sequences for this system are not blends. However, as will be
shown 
below, one of the observed curves can be disentangled into two quite
clearly 
shifted source curves. Why it is so we do not know yet, but it is not 
ruled out that the well-known ``small delay difference'' controversy may be 
connected with this effect 
(see Yonehara~\cite{Yonehara01}; Gil-Merino et al. \cite{GilMerino01}; Goicoechea~\cite{Goicoechea}; Yonehara et 
al.~\cite{Yonehara02}).
Finally,
we 
give a few recommendations for observers on how to plan long monitoring 
programmes for not
fully 
resolved images.

\section{Method}

Here we use a slightly different formalism to introduce the new method 
(cf. original introduction of dispersion spectra in Pelt et 
al.~(\cite{Pelt94,Pelt96})).
In general, the source quasar image in a gravitational lens system is 
split due to 
the gravitational field of the intervening galaxy into multiple images 
$f_1,\dots,f_K$. The source variability $q(t)$ shows itself in the 
measured lightcurves. 
Because of different flight paths the total {\it flight times}
$t_k,k=1,\dots,K$ 
differ. Consequently, the observed luminosities can be described as 
shifted (and
possibly 
magnified) functions of the source variability $f_k(t)=a_k q(t-t_k)$. For 
a fully 
resolved case we will have in total $K$ continuous curves $f_k(t)$. This
somewhat 
oversimplified model ignores microlensing effects (variability of 
the magnification 
coefficients $a_k$ in time) and also other possible distortions.
For each pair of images $f_i,f_j$ we have the corresponding time delay 
$\Delta 
t_{i,j}=t_j-t_i$. From $N (N-1)/2$ delays only $N-1$ can be considered as 
independent. If we know all these independent delays, we can talk about 
a {\it full set} of time delays.

\subsection{Two fully resolved lightcurves}
 
Let us have two unblended images $f_1(t)=a_1 q(t-t_1)$ and $f_2(t)=a_2
q(t-t_2)$. 
We can shift the second curve by a time delay $\Delta t$
and multiply it by an arbitrary magnification ratio $a$ to form a 
difference
$d(t)$ 
\begin{equation}
d(t,\Delta t,a)=f_1(t)-a f_2(t+\Delta t).
\end{equation}
If it happens that $\Delta t = \Delta t_{1,2}=t_2-t_1$ and $a=a_1/a_2$ then
the difference
vanishes, $d(t,\Delta t,a)=0$, and we say that the two curves {\it match} 
each other.
In the case of noisy curves the match cannot be perfect. However, we 
expect that the
dispersion of the difference is minimised for proper parameter values.

Finally, for discrete measurement series, we can hope that for each 
set of
trial
parameter values there are at least some time point pairs that 
can be used to estimate the dispersion. This dispersion is used then as a 
{\it merit function} 
to
compare
different combinations of parameters.
This simple example shows that we can start from continuous curve models
to build (using free parameters) matching pairs and then translate our algorithm
into
language of dispersion spectra. For the details of the method see
Pelt~(\cite{Pelt96}).

\subsection{More than two fully resolved lightcurves}

Sometimes we may have more than two fully resolved images and corresponding 
lightcurves (see Pelt~\cite{Pelt98}). Just for  simplicity let us look at 
a
case 
when we have three images $f_1(t)=a_1 q(t-t_1)$, $f_2(t)=a_2 q(t-t_2)$, and 
$f_3(t)=a_3 q(t-t_3)$. Now we have three amplifications and three time delays 
from which only two are independent. There are also three different
possibilities to 
match curves. As shown in Pelt~(\cite{Pelt98}) it is reasonable to add 
dispersions from all three matches. This allows us to use 
information maximally in the sampled and noisy curves. The number of independent 
variables in 
this case is four~-- two delays and two amplification ratios.

\subsection{A clean image and a blend}

For three images we have basically only one interesting scheme with blended
images. 
In this case we can observe one pure image, say $f_1(t)$ and one blend 
$f(t)=f_2(t)+f_3(t)=a_2 q(t-t_2)+a_3 q(t-t_3)$. These curves cannot be
matched. 
However, we can use a ``clean'' curve $f_1(t)=a_1 q(t-t_1)$ to build an
artificial 
blend curve $g(t)= f_1(t)+ \alpha f_1(t-\Delta_s)$, where $\alpha$ and 
$\Delta_s$
are 
additional free parameters to be determined. In the terms of the source curve we
get 
$g(t)= a_1 q(t-t_1)+ \alpha a_1 q(t-t_1-\Delta_s)$. It is not hard to see that
in a
fortunate case when $\Delta_s=\Delta t_{2,3}$ and $\alpha=a_3/a_2$, the 
curves
$f(t)$ 
and $g(t)$ are shifted and amplified versions of each other and can be matched.
Then we will have three parameters to vary: the two artificial
blend parameters
$\alpha$, $\Delta_{s}$ and the matching parameter $\Delta_l=\Delta
t_{1,2}=t_2-t_1$.
In the matching process we also estimate (for every parameter triple) two additional regression parameters: $a$ and 
$b$, which are discussed briefly in Sect.~\ref{subtraction_of_curves}.
As we see, in addition to subtraction (to match curves), we must also know how
to add 
discretely sampled curves to build artificial blends.

\subsection{Two blends}

If we have four images, then there are many possibilities.
The 
unblended scheme can be treated similarly to the case of three images. The only 
difference is that the combined dispersion has to be estimated, using 
the sum of 6
pairwise 
matches. The number of free parameters is 6 and the search grid can be 
quite large. 
However, it is sometimes possible to reduce the number of free 
magnification ratios.
For the case of one blend and two clean images we can combine clean images 
into an
artificial blend and then compare the combined curve with the measured 
blend curve.

The most interesting case is where we have only two blends, say 
$g_1(t)=f_1(t)+f_2(t)$ and $g_2(t)=f_3(t)+f_4(t)$. It is not possible to build
proper 
matches in this general case, but very often the blend components (in both
blends) 
are of nearly equal luminosity. In this case we can form two artificial blends,
one 
from the first observed blend curve and another one from the second curve. 
Now we can estimate values for input parameters using 
the dispersion of the differences between two artificial blend curves. If 
it happens that the delay we used to build 
the second 
artificial blend is just the delay between $f_1$ and $f_2$ and {\it vice 
versa}, that is,
if the 
delay with which we built the first artificial blend is equal to the delay 
between
$f_3$ 
and $f_4$, then these artificial blends are shifted versions of each other. (The
delay 
between the artificial blends is the third time delay to be varied.)
In a similar way we can proceed further to more complex systems. 
Unfortunately,
the 
parameter space is becoming untreatably large for such systems, although we can
always 
choose some image subsets and then apply one of the simpler schemes.

\subsection{Subtraction of sampled curves}\label{subtraction_of_curves}

Till now we presented possible algorithms in the language of continuous 
lightcurves. The real data is always sampled, and consequently we need to 
reformulate our algorithms for this case.
For every pair of input data tables $t_n,f_n,W_n,n=1,2,\dots, N$ and
$t_m,g_m,W_m, 
m=1,2,\dots,M$ (where $W_n=1/(\Delta f_n)^2$ and $W_m=1/(\Delta g_m)^2$ are 
statistical weights computed from standard errors $\Delta f_n$, $\Delta g_m$
given by the
observer), we can define their ``statistical'' difference or distance between
two 
curves.

To be consistent with actual codes implementing the
proposed methods, we assume that the first curve can be amplified by 
a certain coefficient 
$a$ and it can have a different baseline value $b$.
For a particular set of input parameters 
we can form a table of triples: 
\begin{equation} 
{t_n+t_m \over 2}, (af_n+b-g_m)^2, W_{n,m},
\label{vaheruudud}
\end{equation} 
where $W_{n,m}$ are the statistical weights for every row. 
The actual values for the $a$ and $b$ parameters are to be 
estimated using the least-squares routine, and they are always calculated for 
every set of $\alpha$, $\Delta_{s}$, and $\Delta_l$ in our method.
All the rows in this table are not equally significant. If it 
happens that $t_n=t_m$, 
then we can assign a full weight to the corresponding row. But if the time 
difference between the two 
points is quite large (say larger than a certain pregiven value $\sigma$), 
then comparing the
values for different curves does not make sense. 
Following these 
heuristics we introduce the following {\it downweighting function}: 
\begin{equation}
S_{n,m}=\cases{1-{|t_n-t_m|\over \sigma },& if $|t_n-t_m|\leq\sigma $,\cr
               0,& if $|t_n-t_m|>\sigma$ \cr}. 
\label{sigmad}
\end{equation} 
Finally, the combined 
statistical weights for every row in the table of squared differences 
(Eq.~\ref{vaheruudud}) can 
be written as: 
\begin{equation}
 W_{n,m}=S_{n,m}{W_n W_m \over W_n+a^2 W_m}. 
\label{vahe_ruutude_kaalud}
\end{equation} 
The normalised estimator of the dispersion of the difference between 
the two curves
is now 

\begin{equation}
D^2 =\min_{a,b} {\sum_{n,m} (af_n+b-g_m)^2 
W_{n,m} \over \sum_{n,m} W_{n,m}}, 
\label{distance}
\end{equation} 
and we may call it {\it statistical distance}. From the point of view 
of the parameter estimation scheme, it is also a {\it merit function} to compare
different sets of parameters. 
  
Because one of the parameters we search for, $a$, is included in the 
weight system, the minimisation proceeds 
iteratively. We first fix $a=1$, and compute the weights using this value. 
Then we use a standard weighted least-squares routine to estimate both the
parameters $a$ and $b$. By inserting the estimated $a$ back into 
the weights, we can 
proceed iteratively until convergence is achieved. Fortunately, only a 
small number (about four for 0.1\% precision) of iterations is 
needed.

In some cases the parameter combination that minimises the statistical 
distance can
be unphysical. For particular shifts one curve can approximately match 
the mirror
image of the other, and then the $a$ value can be negative. The distance 
computation
procedure
must take this possibility into account and mark unphysical parameter
combinations. 
It must be said that the described "statistical" subtraction procedure is 
quite 
general. Input data sets can be original data tables,
data with shifted time arguments, or artificial blends computed from input data
by 
adding time-shifted variants of it.

\subsection{Addition of sampled curves}

We start again from two input tables $t_n,f_n,W_n,n=1,2,\dots, N$ and 
$t_m,g_m,W_m, m=1,2,\dots,M$. Now we must compute a discrete analogue to 
the combined curve $f(t) + \alpha g(t)$. Using similar considerations as 
above for the case of subtraction, we can form the triples 

\begin{equation} 
{t_n+t_m\over 2}, f_n+\alpha g_m, W_{n,m},
\label{k6verate_summa} 
\end{equation} 
where the combined weights 

\begin{equation} 
W_{n,m}=S_{n,m}{W_n W_m \over \alpha^2 W_n+ W_m },
\label{summa_kaalud} 
\end{equation} 
consist of appropriately propagated weights and downweighting function.

\begin{figure}
\resizebox{\hsize}{!}{\includegraphics{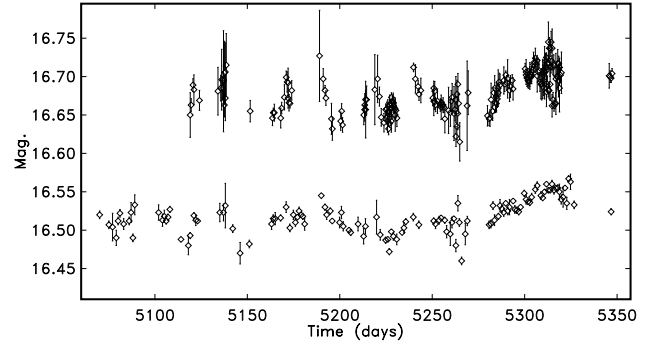}}
  \caption{Combining an original time series with its shifted version. 
Depending on the spacing of time points and the downweighting parameter 
$\sigma$, the resulting series can be sparser (middle part of the series) 
or denser (right part). The combined error estimates are larger than the 
original (given) values.}
  \label{Fig01}
\end{figure}

The total number of selected triples (with non-zero weights) depends on 
the downweighting parameter $\sigma$. In Fig.~\ref{Fig01} we have added an 
original time series (lower part of the figure) and its shifted version to 
form a combined curve (upper part of the figure). In the case of a dense 
sampling, our constructed data set is quite redundant, 
especially for larger values of $\sigma$. If the sampling step is smaller 
than $\sigma$, we may get sparser time series as well. It is very hard to 
choose a proper value for $\sigma$ from purely theoretical considerations. 
However, the proper range of usable values can be established by using 
model or trial calculations.
The sets of triples Eq.~\ref{vaheruudud} or Eq.~\ref{k6verate_summa} can be looked upon as a new input data set for further 
operations.
Combining shifting in time and the adding and subtracting of discrete lightcurves, 
we can now build different discrete models for algorithms described above 
for the continuous case.

\section{Time delays from a clean curve and a single blend} \label{ptk3}

As shown in the previous section, there are a number of possibilities to build 
interesting algorithms, which allow us to search for multiple time delays 
from blended lightcurves.
To be specific, we restrict ourselves to only one particular scheme, where 
we have two observed curves: a clean curve $C$ with a flight time $t_1$ 
and a blend 
$A=A1+A2$, where the flight time for $A1$ is $t_2$ and the flight time for 
$A2$ is $t_3$. We use
the clean 
curve to build artificial blends $B(t)=C(t)+\alpha C(t-\Delta t_{2,3})$ and
compare the
artificial blends with the observed blend $A$. The additional parameter --
the magnification 
factor $\alpha$ -- takes into account the possibility that the two source 
images $A1$
and 
$A2$ have different amplifications.  For every set of trial delays $\Delta
t_{1,2}$, 
$\Delta t_{2,3}$ and the factor $\alpha$ we statistically subtract two 
blends and 
evaluate their distance $D^{2}$.

A global search for the best parameter combination consists of computing the 
statistical distances for a large grid of parameter triads. For some of 
the triads the distance computation can reveal unphysical matches. These 
combinations are discarded and are also marked as special in the 
corresponding multidimensional plots. The best parameter combination 
corresponds to the global minimum of $D^{2}$ and is searched for only 
among physically plausible solutions.

\begin{figure}
\resizebox{\hsize}{!}{\includegraphics{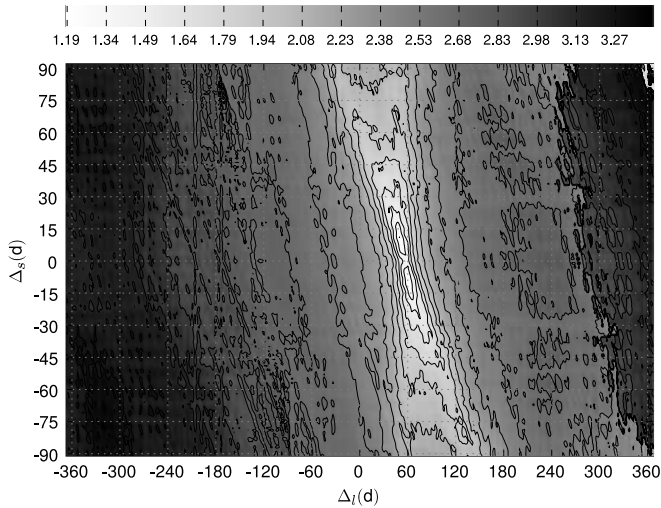}}
  \caption{The two-dimensional grid of merit function values for a 
computer
generated random walk and a blend computed from it. The general minimum
must indicate the true pair of long and short delay values. The plot
demonstrates degeneracy in full-scale computations well~-- there 
is
obvious symmetry between the areas for positive and negative values of
short delays. Values on the colour key represent the $log(D^2)$
and spacing of the contours. The same type of colour key is used in all
two-dimensional plots.}
  \label{dm9}
\end{figure}

There is one interesting aspect in this global search procedure -- it is 
essentially degenerate. The degeneracy comes from the 
fact that the
long and short delays between the blend components can be computed 
differently. The short delay depends on how we assign names to 
hypothetical parts of the blend. In one case the delay is $t_3-t_2$, but 
in another case $t_2-t_3$. And corresponding long delays will be also 
different: $t_3-t_1$ and $t_2-t_1$. This degeneracy results in 
symmetrically placed minima on the grid of the time delays (see for 
instance Fig.~\ref{dm9}). For finite sequences both solutions can give 
slightly different values for the merit function because of the boundary 
effects. Sometimes physical considerations can define the proper order of 
total flight times and then we do not need to compute full grids, but can 
restrict our computations to only one half of them.

It is also important to check how the actual distribution of time moments 
in input data sets influences the statistical weights in the final 
expression for dispersion. Even in the case of the comparison of two pure (one 
of them shifted) lightcurves, it is not ruled out that for a particular 
time delay the observations of one curve occur just in the gaps of the 
other. The well-known controversy on the time delay of the classical 
double quasar QSO 0957+561 was just a result of this kind of accident (see 
for details Press et al.~\cite{Pressa,Pressb}; and Pelt et 
al.~\cite{Pelt94}).
Multidimensional graphs of the parameter dependent sums of weights 
from Eq.~\ref{distance} can reveal regions where there is not sufficient 
information to estimate the parameters of the model.

\section{Tests with simulated data sets} 

Our goal in this paper is to show that the approach described above can be 
useful -- if not generally -- then in many interesting situations. We do 
not have observed data sets for real systems at our disposal that are 
long enough and have been measured for blended systems. Thus we have to 
use computer-generated models. We hope that availability of the new method 
gives an extra motivation for astronomers observing at telescopes with 
modest resolutions to carry out long monitoring programmes for 
gravitational lens systems.

The generation of simulated data is simplified by the fact that model curves 
for different images can be computed from the same source curve. We can 
apply different shifts to time points of a given or generated sampling 
scheme, then interpolate values from the source curve at shifted positions, 
and finally combine the obtained values into blends.
To achieve an appropriate degree of realism, we often used time points and 
weighting sequences from real observations. This allowed us to analyse 
very irregular and inhomogeneous distributions with caps.

\begin{figure}
\resizebox{\hsize}{!}
{\includegraphics{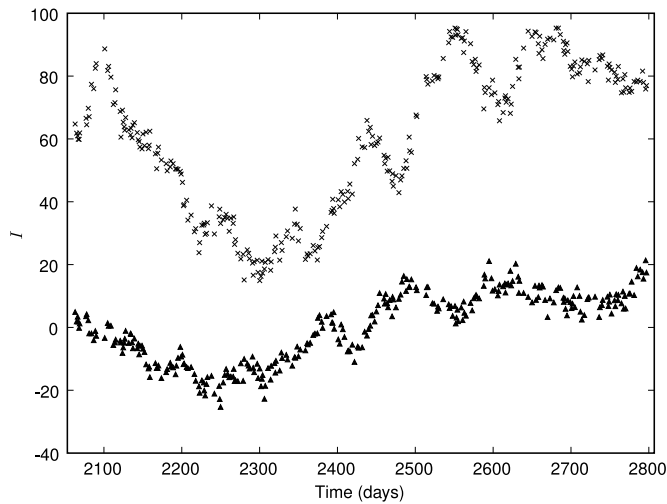}}
 \caption{A basic computer-generated random walk $C$ (lower curve) and a
computer generated blend curve $A$ ($\Delta_{l} = 50.2$, $\Delta_{s} =
10.6$, the amplification parameter $\alpha = 1.3$; we added $5\%$ noise to
both curves and shifted the blend up by $60$ units).}
 \label{c9_a9_koverad}
\end{figure}

In most cases we used a simple random walk procedure to generate the 
source variability curves. The time steps for the curves were selected 
according to two principles: they must be shorter than typical sampling 
intervals and they must be longer than typical photometric integration 
times. A random value of $\pm 1.0$ was assigned cumulatively to each step 
in the intensity scale. If we wanted to model actual observations (especially 
to use their statistical weight systems), then the resulting curves were 
appropriately scaled. One of the generated curves and the blend 
constructed from it is shown in Fig.~\ref{c9_a9_koverad}.

It was interesting to observe that sometimes the generated curve was quite
poor in features (minima and maxima, etc.). In these cases we discarded 
them. There is a similar effect when dealing with actual
lens systems. A quasar can be ``quiet'' for a long time and its photometry
is not sufficient for time delay estimation.

\subsection{Random sampling}

\begin{figure}
\resizebox{\hsize}{!}{\includegraphics{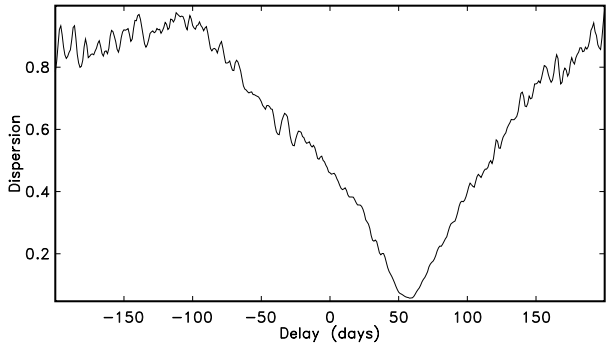}}
 \caption{A classical dispersion spectrum computed for the computer 
generated curve $C$ and the artificial blend $A$. It reveals a shift 
$\Delta t \approx 59$ days. However, the blend was generated by using a 
long delay value $50.2$ days and a short delay $10.6$ days. Consequently, 
blending can mask proper time delay values. The fully resolved case is 
shown in Fig.~\ref{dm9}.}
 \label{disp_spekt}
\end{figure}

We started our analysis with a simple sampling scheme where the initial
time points were generated by using random step sizes from the interval
$[0.2,1.8]$ days. 
The generated time points were then used to read off (using appropriate time
shifts and linear interpolation) three
different data sequences for the images $C$, $A1$, and $A2$. From the two 
last images we formed the blend
$A=A1+\alpha A2$
with a fixed amplification ratio $\alpha$. In a particular example 
displayed in
Figs.~\ref{dm9}~-~\ref{disp_spekt}, 
the following parameters were used: the ``long'' time delay between $C$ 
and $A1$ was
$\Delta_{l}=50.2$, 
the ``short'' delay between the blend components was $\Delta_{s}=10.6$, and 
the amplification
parameter $\alpha=1.3$. To take into account daylight and randomly 
changing observational conditions, part of the time points were discarded. 
A typical run of our scheme is given in Fig.~\ref{c9_a9_koverad}. From the
initially generated $734$ sample points, only $306$ were left 
in the time series. Finally, we added a random 5\% Gaussian noise component 
to the curves $C$ and $A$. 

The two resulting model curves $C$ and $A$ can be used as input for a standard 
time delay estimation procedure. As seen from Fig.~\ref{disp_spekt}, in 
this example there is a quite
well pronounced global minimum in the dispersion curve, but its position
is at $\Delta t \approx 59$ days. Blending can move dispersion minima
from one place to another.

To recover both delays, we need to apply our new method with artificial 
blends. First we form a three-parameter search grid $[-360,360:1.0]\times 
[-90,90:1.0]\times[0.6,1.6:0.1]$\footnote{ Here and below we use a 
systematic notation for search grids. Inside the square brackets we give 
the minimum and maximum values for the parameter in question, followed by 
the grid step. } and compute the merit function $D^2$ values for each grid 
point. The value for downweighting parameter $\sigma$ was taken at
$2.5$ days. (See Sect.~\ref{discussion} for discussion on choosing the 
value of parameter $\sigma$.) The parameter combination for the minimum 
value of the merit function is then selected as the solution. In the 
current example the best triple occurred at $\Delta_l = 50$ days, 
$\Delta_s = 11$ days, and $\alpha = 1.1$. The two-dimensional slice 
at $\alpha=1.1$ of the search grid is given in Fig.~\ref{dm9}. In this 
plot we can clearly see the degenerate character of our procedure. 
Depending on the ordering of the blend components, we can attain the 
deepest minima in two symmetrically placed areas. Formally, both global 
minima are of equal value, but typically the actual merit function values 
for both minima differ slightly. If we do not have any additional 
information (say from deep lens system images) we can just formally select 
the strongest minimum.
Another interesting point of the current example is the fact that our 
algorithm did not exactly recover the amplification ratio parameter (we 
found 1.1 instead of 1.3). This is quite typical -- for every particular 
pair of delay values the merit function dependence on the parameter 
$\alpha$ is quite weak and the corresponding curve has a wide minimum 
around the correct value.

For high quality data (with small errors) we can look for a final solution 
with higher precision. For each strong minima found during the rough 
analysis, we can build refined local parameter grids in the vicinities of 
the preliminary solutions. An example of such local refinement is given in 
the next subsection.

\subsection{Real sampling}\label{real_sampling}

\begin{figure}
\resizebox{\hsize}{!}{\includegraphics{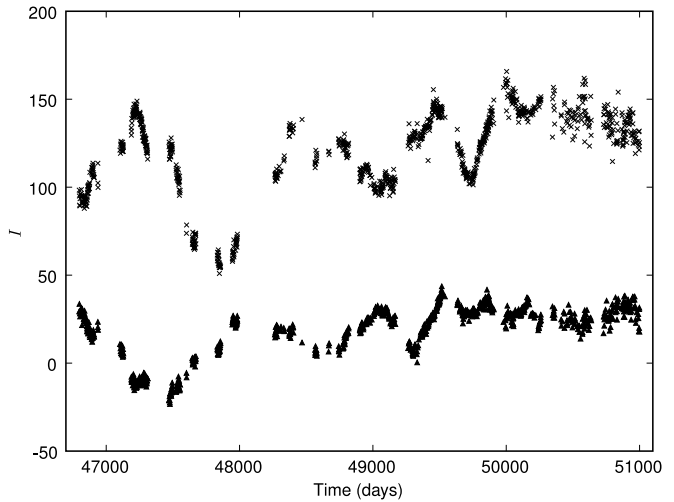}}
\caption{A random walk $C$ (lower curve) and a blend $A$
(upward-shifted curve) with real sampling.}
\label{043_koverad}
\end{figure}

To test our method under real sampling conditions, the time points from 
the observational data by 
Schild\footnote{http://cfa-www.harvard.edu/\~{}rschild/fulldata2.txt} were 
used to sample the $C$ and $A$ curves built from a computer-generated 
random walk pattern. The standard errors given by Schild were scaled 
according to the ratio of the full amplitudes of the real and simulated 
data. Next these scaled errors were used as standard deviations for 
Gaussian noise components, which were added to each point of simulated 
data. The resulting curves for one particular run are shown in Fig. 
\ref{043_koverad}.

\begin{figure}
\resizebox{\hsize}{!}{\includegraphics{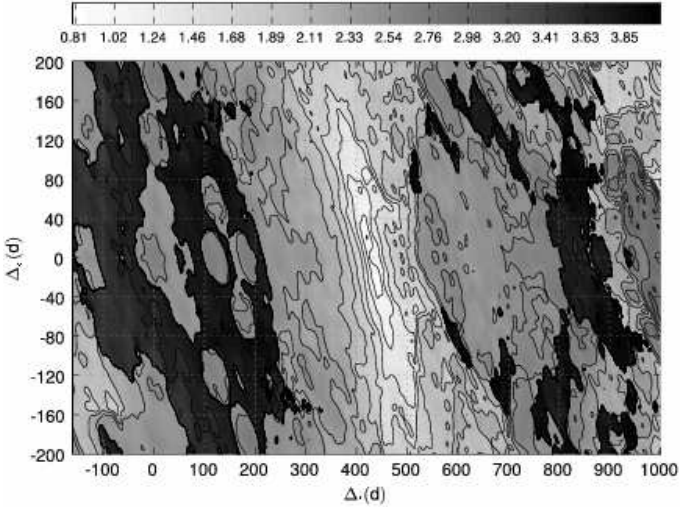}}
\caption{Merit function values for simulated data and real sampling.}
  \label{dm043}
\end{figure}

The values of parameters used in this simulation were: $\Delta_{l}=420.15$ 
days, $\Delta_{s}=20.21$ days, $\alpha=0.8$. We selected a 4202 day 
time-interval from Schild's observations and the number of data points in 
the input table was 1032.
A crude search grid $[-150,1000:1.0]\times 
[-200,200:1.0]\times[0.6,1.6:0.1]$ was used to estimate the amplification 
parameter $\alpha$. The grid slice with the best value $\alpha = 0.9$ was 
then used to refine other two parameters. (Here and in the next 
section the value for the downweighting parameter $\sigma$ was taken at $7$ 
days.) Finally we got the estimates for the delays $\Delta_{l}=420.2$ 
days, $\Delta_{s}=20.0$ days. The plot of merit function values in the 
slice $[-150,1000:0.1]\times [-200,200:0.1]$ is shown in Fig.~\ref{dm043}. 
Again, we can see the two symmetrically placed minima.

\begin{figure}
\resizebox{\hsize}{!}{\includegraphics{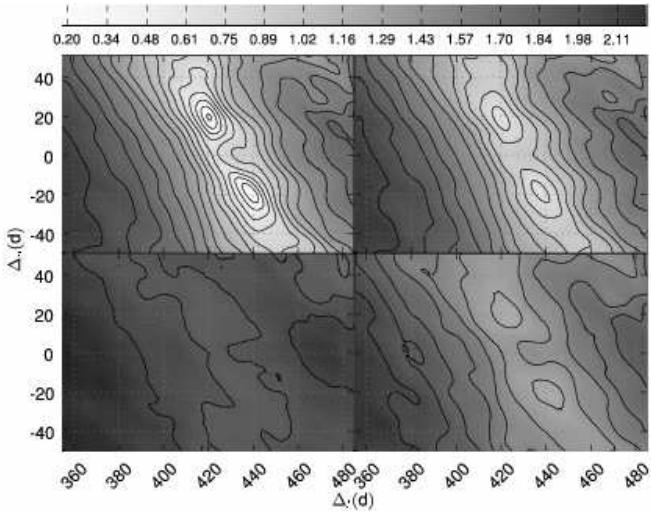}}
 \caption{Vanishing of the characteristic minima of the merit function 
values due to observational errors. The noise levels are (clockwise from 
upper left): 0\%, 2\%, 5\%, and 10\%.}
 \label{dm_myra_tase}
\end{figure}

To see how observational errors would affect the results, we 
performed an additional experiment in the vicinity of the obtained 
solution.
We added gradually varying levels of Gaussian noise to the simulated data 
and evaluated merit function values for the slice $[360,480:0.1]\times 
[-40,40:0.1]$ ($\alpha = 0.9$). The corresponding plots for the four noise 
levels are given in Fig.~\ref{dm_myra_tase}. It is clearly seen how the 
minima are smeared out when the noise level rises. From 
Table~\ref{myratabel} we can see how the global solution behaves when noise 
is added. As was mentioned above, sometimes the solution can jump to its 
mirror place (and in this case we essentially do not lose it). For higher 
noise values we can completely lose the correct solution.
When planning for a long time photometric monitoring it is reasonable to 
establish the appropriate levels of measurement precision. 
For that purpose similar model calculations as carried out in this section 
can be useful.

\begin{table}[htb]
\centering
\caption{\label{myratabel}Location of global minima depending on added 
Gaussian noise.}
\begin{tabular}{c  c c}	   	   \hline \hline
Noise	&$\Delta_{l}$	&$\Delta_{s}$	\\
(percent)	  					\\ \hline	
0 	&420.6	& 19.0 \\
1	&420.5	& 19.9 \\
2	&419.9	& 19.1 \\
3	&438.5	&$-19.0$\\
4	&423.4	& 20.0  \\
5	&419.5	& 26.9  \\
6	&417.5	& 80.9  \\
10	&480.0	& 46.9  \\
15	&499.4	& 80.9  \\
20	&481.4	& 43.9  \\ \hline
\end{tabular}
\end{table}

\subsection{Error estimation} \label{ptk42}

To test statistical properties of the new method we used Monte Carlo type 
calculations. We added appropriately scaled Gaussian noise components to the noise-free model curves from the previous 
section, so that the 
expected signal-to-noise ratios were the same as those for Schild's data.  
Using Estonian GRID\footnote{See http://grid.eenet.ee/en/} resources, we 
repeated this 3500 times and stored the obtained optimal parameter 
triples. From this resulting table we calculated the average values and 
standard deviations: $\Delta_{l}=419.6\pm 0.8$ and $\Delta_{s}=20.14\pm 
1.22$ days, $\alpha=0.94\pm 0.10$. The resulting biases $419.6 - 420.15$, 
$20.14-20.21$ and standard errors for this particular experiment setup are 
reasonably small.

\section{Results for a real system} 

\begin{figure}
\resizebox{\hsize}{!}{\includegraphics{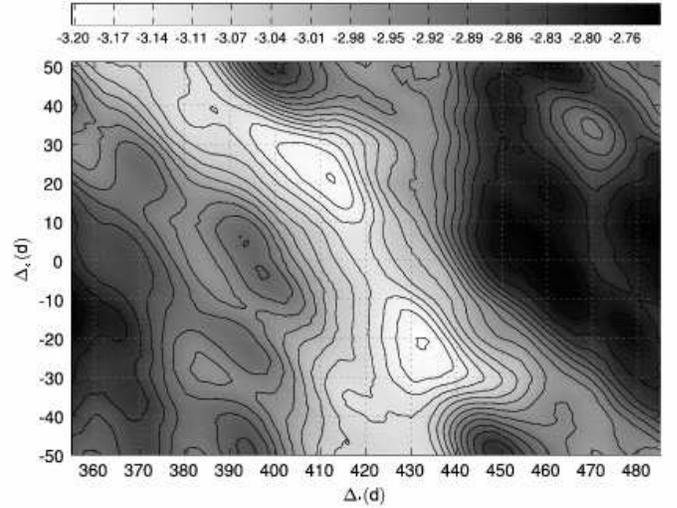}}
  \caption{Merit function values for the actual double quasar data. It 
differs significantly from Fig.~\ref{dm_schild0}.}
  \label{dm_paris_schild}
\end{figure}

To evaluate the new method in a more realistic context, we used the master 
data set for the double quasar QSO 0957+561 A,B kindly provided by Rudy 
Schild (6806 days, 1233 time points, R-band optical CCD 
photometry). As far as we know, the components of the system itself cannot
be considered as blends and consequently we used this data set as a 
model for time point spacing and observational error distribution for a 
long and realistic monitoring programme, as was discussed in Sect.~\ref{real_sampling}.
In the course of experimentation we also performed 
some calculations with the full Schild data set and got unexpected 
results. Assuming that $B$ is a blend, we indeed got a distribution 
characteristic to the blended case, shown in Fig.~\ref{dm_paris_schild}. 
The estimates for the time delays in real data are $\Delta_{l}=412, 
\Delta_{s}=22$ days. The amplification factor was held fixed at $\alpha 
\equiv1.0$. ($\alpha \approx1.0$ is the expected value for short 
$\Delta_{s}$). In the current section, the value for the downweighting 
parameter $\sigma$ was taken $10$ days.

\begin{figure} 
\resizebox{\hsize}{!}{\includegraphics{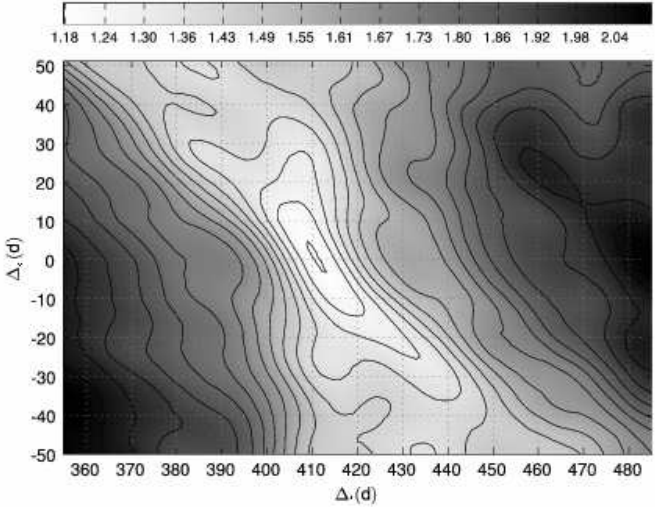}}
  \caption{Merit function values for a random walk and its shifted version 
($\Delta_{l} = 412$ days). Timepoints and standard errors are from Rudy 
Schild's monitoring programme.}
  \label{dm_schild0} 
\end{figure}

To convince ourselves that the
symmetric minima in Fig.~\ref{dm_paris_schild} are not caused
by boundary effects of our computational algorithm, we performed some
additional tests.
First we used the real time moments and error estimates from the same data 
set and built a pair of artificial curves with a given single time delay 
between the two curves $\Delta_{l} = 412$. Then we disturbed both curves 
by appropriately scaled random errors. The resulting two-dimensional slice 
of the merit function is shown in Fig.~\ref{dm_schild0}. It is well seen 
that there is one unique global minimum near the
true delay value, indicating that we do not have a blend here, the delay applied is recovered and the estimated short shift value 
(if we assume that the B curve is a blend) is zero. Consequently, our 
method does not generate symmetrically placed minima just as an artifact 
of the procedure.

\begin{figure}
\resizebox{\hsize}{!}{\includegraphics{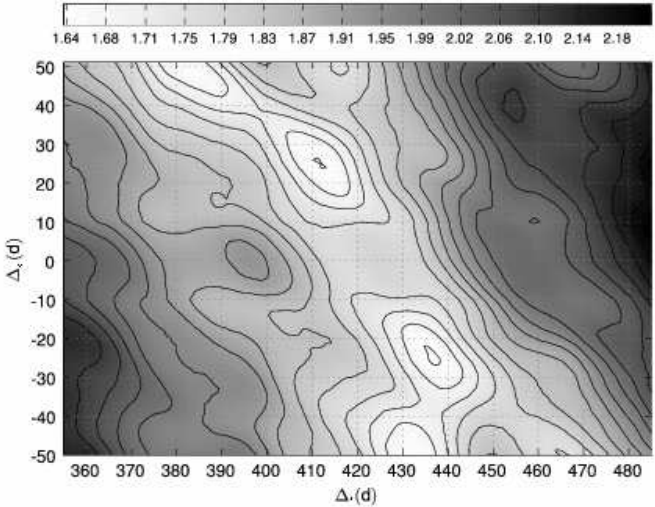}}
  \caption{Merit function values for a computer generated random walk and 
the blend computed from it using the parameters found from the real 
observational data.}
  \label{dm_real2sim} 
\end{figure}

Finally we built an artificial blended model with the long and short 
delays found from the real curves $A$ and $B$. The resulting grid for the 
optimal amplification parameter is shown in Fig.~\ref{dm_real2sim}. From 
the last simulation we found $\Delta_{l}=412$, $\Delta_{s}=25$ days, 
indicating that the time delay values found from the real data are real. 
The three-day long estimation error in short shift characterises the 
precision of the algorithm at the given level of observational accuracy.

\begin{figure}
\resizebox{\hsize}{!}{\includegraphics{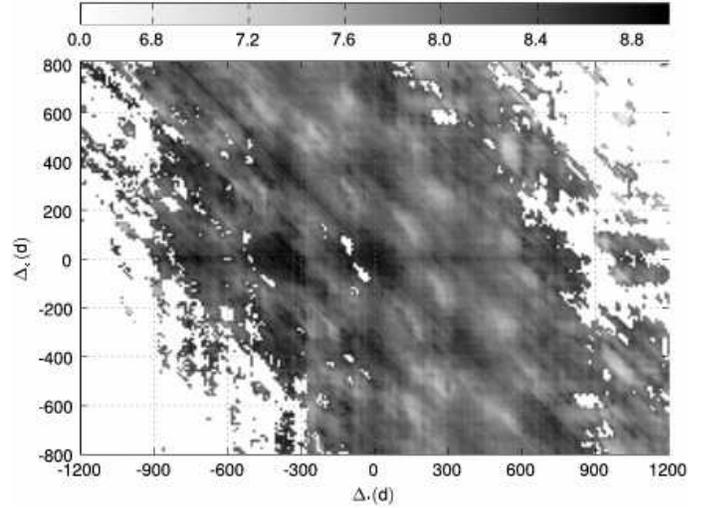}}
 \caption{The expected sums of weights for the time delay space of 
Schild's data. Zero values represent areas with unphysical negative 
parameter $a$ values as described in the text.} 
 \label{kaalupilt}
\end{figure}

To check how statistically stable the merit function values are for 
different parameter combinations, we calculated expected sums of weights 
for the time delay grid of Schild data. As it is seen from 
Fig.~\ref{kaalupilt}, $\Delta_{l}=412$ and $\Delta_{s}=22$ days fall into 
the region of higher weights and should be considered a reliable result.
It is currently very difficult to tell why the $B$ curve of the classical 
double quasar behaves as a blend. It is known that there is something 
wrong with the estimated time delays and magnification ratios (if 
optical data is compared with radio data). The peculiar form of 
microlensing proposed in Press~(\cite{Press98}) can solve the problem of 
magnification ratios. However, it is hard to expect that the spacing of 
microlensing events in time can mimic a proper blend. In another 
development, Goicoechea~(\cite{Goicoechea}) singles out the different 
features in the double quasar lightcurves which give different values for 
time delays. As a possible explanation he uses a quasar model with 
spatially distant flares, as
discussed also in Yonehara~(\cite{Yonehara01}) and Yonehara et al.~(\cite{Yonehara02}).
Similar and even more radical ideas can be found 
in a recent work by Schild~(\cite{Schild05}).
Our computations show that not only single events, but the full $B$ curve 
of the system can be decomposed into a sum of two similar and shifted 
curves. What theoretical interpretation can be given to this phenomenon 
remains an open question.

\section{Discussion}\label{discussion} 

Above, we tried to demonstrate that a very interesting data 
processing method exists, which allows us to estimate integral time-delay systems for 
blended (not fully resolved) lightcurves. There are not yet enough real 
observed sequences to evaluate the full potential of the new method, but 
our 
computations allow us to get at least a preliminary idea about 
the applicability of 
such algorithms.
When planning a new monitoring programme we suggest to consider 
the following
aspects:
\begin{itemize}
\item The total time base of observations must be determined from 
the expected
length of the longest time delays. We cannot sensibly find longer time delays from
the observed time 
series, than about half of the length of the time series itself.
\item Sampling should be dense enough to match the characteristic
periods of variability of the source quasar. In more precise terms -- 
the sums of weights for every parameter combination to be compared 
must be large enough to avoid statistically unstable merit function values.
\item Sampling determines the shortest possible delay we are able to find 
from the particular data.
If the downweighting parameter $\sigma$ (see formula \ref{sigmad}) is too 
small
for a given sampling, we will have too few pairs in the calculation of the 
merit
function and the map of $D^2$ will be poor due to noise and boundary 
effects. On the other hand, enlargement of $\sigma$ is limited because of 
the smoothing effect of this parameter~-- using larger $\sigma$ reduces 
the
possibility of finding shorter time delays. As a rule of thumb $\sigma$
should be kept equal to or smaller than half of smallest possible time 
delay
$\Delta_{s}$ we are trying to find. For sound statistics, we should have on
the average at least 3--5 pairs for every observed time point when combining
and
subtracting the time series. In practice it would be useful to carry on
computations with varying values of $\sigma$ to check statistical 
stability and robustness. 
See for instance Pelt~(\cite{Pelt96}) where such an analysis was used for 
a simple case of delay estimation. 
\item A sufficiently dense and truly random distribution of the time 
points is best for the method described above. The strongest 
source of uncertainty for the described algorithms is periodicity of data
gaps. Unfortunately, ground-based astronomical observations tend   
to be of this kind. To lessen the effect of gaps, combined observations 
from
different
sites can be used.
\item As we saw in Sect.~\ref{real_sampling}, observational errors 
for typical experimental setups must be kept under 
5\% of the amplitude of lightcurve. 
\item As a sanity check, it is worth to compute
merit function values for a larger parameter grid. Then the
overall pattern of 
symmetrically shaped and mirrored minima allow us to get the general 
impression of
the
validity of our solutions.
\end{itemize}

The software modules we have developed can be used to model situations 
that can occur in 
real long-time monitoring programmes. By varying model parameters we can 
estimate sufficient durations for observational sessions and also the 
accuracy 
of 
observation needed. Unfortunately, even accurately planned sessions can 
result in a failure because the source quasars themselves can show persistent 
stationarity or the time series observed can be contaminated by strong 
microlensing.
It is not ruled out that the proposed method can be used in absolutely 
different 
contexts. One of the applications we are presently considering is 
disentangling echo components
in certain high energy events of cataclysmic stars.

\section{Conclusion}

A method of finding time delays from photometry of unresolved
gravitationally lensed quasar images has been developed and tested using 
simulated and real data sets. 
It can be used to analyse already existing data sets and 
also to
plan new observations.
We encourage observers to run long time 
monitoring programmes of blended images to reveal time delays 
from 
obtained data.

\begin{acknowledgements}
      We are indebted to the anonymous referee, E. Saar, and T. Viik for many valuable 
comments.
      Part of this work was supported by the Estonian Science Foundation
      grant No. 6813. 
      For time consuming computations, Estonian Grid resources were used.	
\end{acknowledgements}


\begin{thebibliography}{}
  \bibitem[1997]{Barkana} Barkana R., 1997, ApJ 489, 21 
  \bibitem[2001]{Burud} Burud I., Magain P., Sohy S., Hjorth J., 2001, A\&A 380,
805
  \bibitem[2004]{Chartas} Chartas G., Dai X., Garmire G.P., in {\it Carnegie
Observatories Astrophysics Series, Vol.2: 
                          Measuring and Modeling the Universe}, ed. W.L.
Freedman, 2004
  \bibitem[2003]{Dai} Dai X., Chartas G., Agiol E., Bautz M.W., Garmire G.P.,
2003, ApJ 589, 100
  \bibitem[1999]{Fassnachta} Fassnacht C.D., Pearson T.J., Readhead A.C., et al.,
1999, ApJ 527, 498 
  \bibitem[2002]{Fassnachtb} Fassnacht C. D., Xanthopoulos E., Koopmans L. V.
E., Rusin D., 2002, ApJ 581, 823
  \bibitem[1996]{Geiger} Geiger B., Schneider P., 1996, MNRAS 282, 530 
  \bibitem[2001]{GilMerino01} Gil-Merino R., Goicoechea L. J., Serra-Ricart M.,
et al., 2001, MNRAS 322, 397
  \bibitem[2002]{GilMerino} Gil-Merino R., Wisotzki L., Wambsganss J., 2002,
A\&A 381, 428
  \bibitem[2002]{Goicoechea} Goicoechea L.J., 2002, MNRAS 334, 905
  \bibitem[1992]{Hjorth} Hjorth P.G., Villemoes L.F., Teuber J.,
Florentin-Nielsen R., 1992, A\&A 255, L20 
  \bibitem[2006]{Kochanek} Kochanek C. S., Mochejska B., Morgan N. D., Stanek K.
Z., 2006, ApJ 637, L73
  \bibitem[1997]{Kundic} Kundi\'c T., Turner E.L., Colley W.N., et al., 1997,
ApJ 482, 75 
  \bibitem[2004]{Morgan} Morgan N.D., Caldwell J.A.R., Schechter P.L., Dressler
A., Egami E.,  Rix H.-W., 2004, AJ 127, 2617 
  \bibitem[1994]{Pelt94} Pelt J., Hoff W., Kayser R., Refsdal S., Schramm T.,
1994,A\&A 286, 775 
  \bibitem[1996]{Pelt96} Pelt J., Kayser R., Refsdal S., Schramm T., 1996, A\&A
305, 97 
  \bibitem[1998]{Pelt98} Pelt J., Hjorth J., Refsdal S., Schild R., Stabell R.,
1998, A\&A 337, 681
  \bibitem[2005]{Pelt05} Pelt J., 2005, in e-Proceedings of the GLQ Workshop
(http://grupos.unican.es/glendama/e-Proc.htm), C5 
  \bibitem[1997]{Pijpers} Pijpers F.P., 1997, MNRAS 289, 933 
  \bibitem[1992a]{Pressa} Press W.H., Rybicki G.B., Hewitt J.N., 1992a, ApJ 385,
404
  \bibitem[1992b]{Pressb} Press W.H., Rybicki G.B., Hewitt J.N., 1992b, ApJ 385,
416
  \bibitem[1998]{Press98} Press W.H., Rybicki G.B., 1998, ApJ 507, 108 
  \bibitem[1997]{Schild} Schild R., Thomson D.J., 1997, AJ 113, 130
  \bibitem[2005]{Schild05} Schild R., 2005, AJ 129, 1225
  \bibitem[1999]{Yonehara01} Yonehara A., 1999, ApJ 519, L31
  \bibitem[2003]{Yonehara02} Yonehara A., Mineshige S., Takei Y., Chartas G., Turner E.L., 2003, ApJ 594, 107

\end{thebibliography}
\end{document}